\documentstyle[aps,epsf,preprint]{revtex}
\begin{document}
\newcommand{\kv}{{\bf k}}
\newcommand{\qv}{{\bf q}}
\newcommand{\pv}{{\bf p}}
\newcommand{\Pv}{{\bf P}}
\newcommand{\xv}{{\bf x}}
\newcommand{\zv}{{\bf z}}
\newcommand{\Jv}{{\bf J}}
\newcommand{\Av}{{\bf A}}
\newcommand{\Ev}{{\bf E}}
\newcommand{\Qv}{{\bf Q}}
\newcommand{\tz}{t_{0}}
\newcommand{\tm}{T_{\rm 0}}
\newcommand{\DOS}{N(0)}
\newcommand{\xvz}{{\xv}_{0}}
\newcommand{\tilM}{{\tilde{\Omega}}}
\newcommand{\tilt}{{\tilde{t}}}
\newcommand{\Jzero}{{J^{\rm (0)}}}
\newcommand{\Ja}{{J^{\rm (A)}}}
\newcommand{\arma}{{\rm a}}
\newcommand{\armb}{{\rm b}}
\newcommand{\nimp}{n_{\rm i}}
\newcommand{\vi}{v_{\rm i}}
\newcommand{\Ap}{A_{p}}
\newcommand{\App}{A_{p'}}
\newcommand{\tauphi}{\tau_{\varphi}}
\newcommand{\phase}{\phi}
\title{
Transient electric current through an Aharonov-Bohm ring
after switching of a Two-Level-System
} 
\author{Gen Tatara\\
Graduate School of Science, Osaka University, Toyonaka, Osaka 560-0043, 
Japan
}
\sloppy
\maketitle
\begin{abstract}
    Response of the electronic current through an Aharonov-Bohm ring 
    after a two-level-system is switched on
    is calculated perturbatively by use of 
    non-equilibrium Green function. 
    In the ballistic case the amplitude of the Aharonov-Bohm 
    oscillation is shown to decay to a new equilibrium value
    due to scattering into other electronic states. 
    Relaxation of Altshuler-Aronov-Spivak oscillation in diffusive 
    case due to dephasing effect is also calculated. 
    The time scale of the relaxation is determined by characteristic 
    relaxation times of the system and the splitting of two-level-system.
    Oscillation phases are not affected.
    Future experimental studies of current response may give us direct 
    information on characteristic times of mesoscopic systems. 
\end{abstract}

\section{Introduction}

Decoherence (or dephasing) caused by external perturbations is an  
important problem of quantum systems. 
Within equilibrium statistical mechanics, a convenient formula for 
estimating the dissipation by the 
environment was presented by Caldeira and Leggett\cite{Caldeira81}.
There decoherence was treated as an interaction non-local in the
imaginary time. 
The formula was shown to be useful in considering macroscopic quantum 
phenomena\cite{Caldeira81}, in which the tunneling rate was calculated  
only as a static quantity.

Effects of decoherence on electron systems was studied in the 80's 
in the context of weak localization  (e.g., decoherence by phonons and 
electron-electron interaction)\cite{Bergmann84,Lee85}. 
Decoherence gives rise to a mass of 
electron-electron propagator (cooperon), which governs 
magnetoresistance.  
Decoherence time due to electron-electron interaction was calculated 
by solving the cooperon equation\cite{Altshuler82} and 
as a mass of the cooperon\cite{Fukuyama83}.
Later it was demonstrated that this dephasing time is equivalent to 
the time defined in an intuitive way from a decay of the overlap of 
wave function\cite{Stern90,Imry97}.
One should note, however, that this definition does not always work 
as we see below.

Recently decoherence by a quantum two level system (TLS) has been 
theoretically studied\cite{Imry99,Ahn01,Aleiner01}.
In these works temperature dependence of dephasing time, 
$\tau_{\varphi}$, is calculated, motivated by experimental 
finding of the saturation of dephasing time as $T\rightarrow0$ 
in disordered metal\cite{Mohanty97}. The mechanism of saturation 
appears still controversial.

For studies of decoherence, recent mesoscopic systems are 
suitable, since decoherence can be detected in a controlled 
manner. 
A direct way to study decoherence
is to use the interference of two different 
paths in a small ring. 
The interference leads to an oscillation of 
conductance as a function of external magnetic flux through the ring 
(Aharonov-Bohm (AB)\cite{Aharonov59} and Altshuler-Aronov-Spivak 
(AAS)\cite{Altshuler81}oscillations\cite{Aronov87}). 
The oscillation pattern changes if a perturbation causes scattering 
or dephasing. 
The first direct measurement of the effect of the phase due to 
transport through a quantum dot was carried out by use of AB effect 
by Yacoby et. al.\cite{Yacoby95}.
Further studies revealed rigidity of the phase, which is
consequence of time-reversal 
symmetry\cite{Yacoby96,Hackenbroich96,Schuster97} .
The amplitude and phase of AB oscillation was calculated in the 
presence of a dot driven by a AC field in Ref. \cite{Jauho98}.
The effect of a time varying potential on conductance of a ring was 
calculated in Ref. \cite{Yakubo00}.

Recently AB effect in the ballistic case has been experimentally 
investigated\cite{Pedersen00,Hansen01}.
It was argued that the temperature dependence of the AB amplitude 
indicates dephasing rate proportional to $T^{-1}$\cite{Hansen01}.
This behavior was discussed to be consistent with theoretical 
estimate of the dephasing due to charge fluctuation taking account of 
the existence of the leads\cite{Seelig01}.
However the argument given in Ref. \cite{Hansen01} might be too naive 
because theoretically the role of dephasing on AB effect in the 
ballistic case is not obvious, since
dephasing is incorporated by 
cooperon, which exists only in the dirty case.
A possible dephasing's effect on ballistic current may be 
by changing the spectral function\cite{Murphy95}.

The aim of this paper is to study the response of the current through 
a narrow ring with a magnetic flux after a time-dependent environment 
is switched on.
By use of measurement of electronic 
properties with high (THz) time-resolution\cite{Beard00}, 
observation of such current response and time-resolved dephasing 
process would be possible. 
The current response may provide us direct information on microscopic 
relaxation times (elastic ($\tau$) and inelastic lifetime 
($\tauphi$)) and properties of the perturbation source.
As the environment we take TLS. 
The transient current at low temperatures 
is calculated diagrammatically using 
non-equilibrium green function\cite{Keldish64,Haug98,Rammer86}.
Coupling to TLS is included to the second order and a linear response 
with respect to the probe electronic field is considered. 
AB current is calculated in the ballistic case treating the arm of the
ring as one-dimensional.
(The response of AB current as sample dependent fluctuation
in a dirty case would be similar to that of AAS current.) 
A generic expression of the AB response is obtained in 
terms of the correlation functions of the perturbation source. 
It is shown that only the amplitude of the AB oscillation is 
affected, consistent with the 
phase rigidity\cite{Yacoby96,Hackenbroich96}.
The reduction of the amplitude is shown to be simply due to the 
scattering into other electron states and is not interpreted as 
dephasing.
The overlap of the wave function with the initial state exhibits a 
decay after 
TLS is switched on, but this is nothing to do with dephasing.
This is in contrast to the decay caused by
electron-electron interaction in a disordered case\cite{Stern90}.
Theoretically this difference is natural since dephasing in the 
strict sense can not be described by one-particle propagation (
a Green function with an elastic lifetime). The effect is 
incorporated only when we take into account the particle-particle 
ladder (cooperon), which 
represents the interference between a path and 
the reversed path  in the presence of elastic impurity scattering.
Physically interaction with a single TLS itself gives a 
definite phase factor and cannot cause dephasing in the ballistic case.
To cause dephasing, some randomness such as impurities 
is needed to give uncertainty to the phase due to the interaction. 
If there are many TLSs with different energy, dephasing would appear 
even in the ballistic case.

In the calculation of AAS current (\S\ref{SECAAS}), 
dephasing time $\tauphi$ is included 
phenomenologically (we do not care about the origin here).
The lowest order contribution we calculate corresponds to the 
correction to $\tauphi$ by TLS.
The calculation of response of the AAS current is very complicated and 
hence we show the leading term only.

Effect of oscillating external field is briefly discussed in 
\S\ref{SECAC}.

\section{Formulation}
\begin{figure}[bthp]
\begin{center}
\epsfxsize=10cm              
\epsfbox{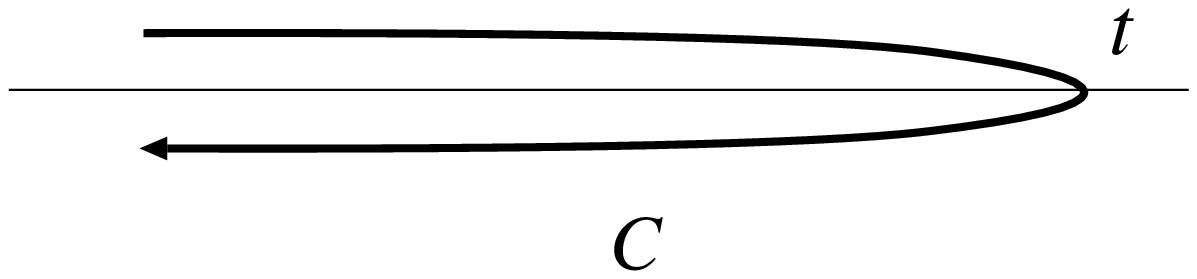}
\end{center}
\caption{Path in the complex time plane. $t$ is the time of 
measurement.
\label{FIGAB_c}}
\end{figure}

\begin{figure}[bthp]
\begin{center}
\epsfxsize=10cm              
\epsfbox{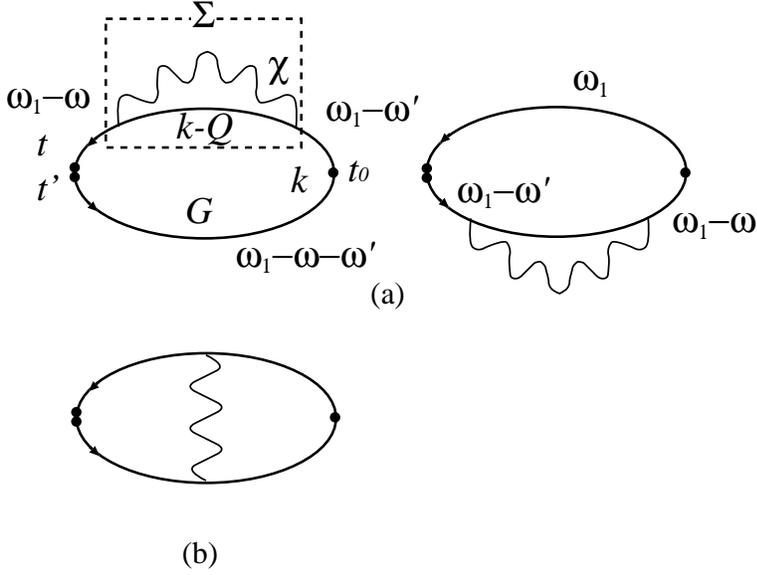}
\end{center}
\caption{Second order contribution to $Q$. (a): Self-energy type. (b): 
Vertex correction type, which vanishes since interaction vertex $V$ 
does not depend on the momentum transfer.  \label{FIGAB_se}}
\end{figure}

\begin{figure}[bthp]
\begin{center}
\epsfxsize=10cm              
\epsfbox{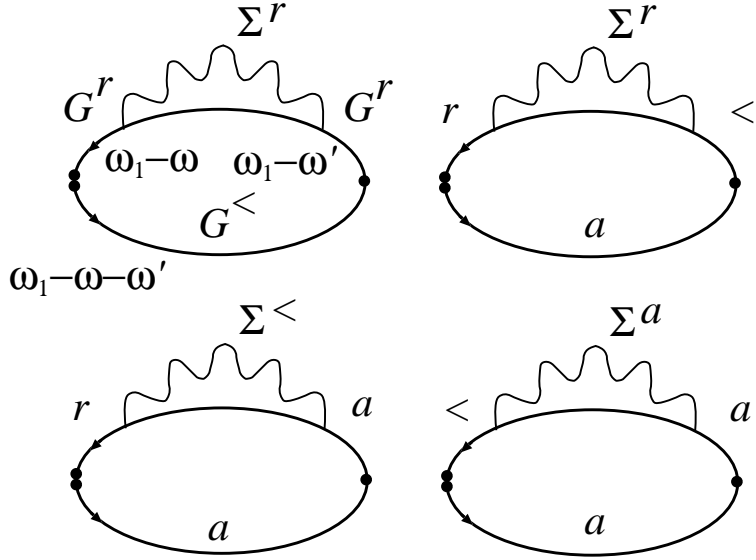}
\end{center}
\caption{Decomposition of $Q^<$ into retarded 
($r$), advanced ($a$) and lesser components ($<$). \label{FIGAB_seKL}}
\end{figure}

The hamiltonian we consider is $H=H_{\rm e}+H_{\rm TLS}+H'$, where
$H_{\rm e}=\sum_{\kv} \epsilon_{\kv} c^\dagger_{\kv} c_{\kv}
+H_{\rm imp}$ is the 
electron part ($\epsilon_{\kv}=\kv^2/(2m)-\epsilon_{F}$, 
$\epsilon_{F}$ being Fermi energy) and $H_{\rm 
imp}\equiv \vi\sum_{\kv\kv'} c^\dagger_{\kv'} c_{\kv}$ represents impurity 
scattering($\vi$ is coupling constant).
$H_{\rm TLS}$ is the hamiltonian of the TLS, which we describe later. 
The coupling between the electron and TLS is
\begin{equation}
    H'(t)=\sum_{\kv\Qv} V(t) c^\dagger_{\kv+\Qv}c_{\kv}
    \label{eq:Hintdef}
\end{equation}
where $V(t)$ is an operator of TLS, which is time-dependent. 
In this paper $V$ is treated as independent of momentum transfer, 
$\Qv$, assuming a small perturbed area. 
We consider an electronic field ($\Ev$) applied on a lead with the 
frequency of $\omega$. The vector potential $\Av$ is then written as 
$\Av(t)=\frac{1}{i\omega}\Ev e^{-i\omega t}$.
We consider a limit of $\omega\rightarrow0$ and $\Ev\parallel z$. 
The electronic current in linear response is given as 
$J=\Jzero+\Ja$, where
\begin{eqnarray}
   \Jzero(\xv,t) & = & \frac{1}{V}\frac{E_{z}}{\omega}
               \left(\frac{e}{2m}\right)^2 
   (\nabla_{\xv}-\nabla_{\xv'})_{z} (\nabla_{\xvz}-\nabla_{\xvz'})_{z}
      Q^{<}_{\xv\xv',\xvz\xvz'}
  (t,t',\omega )|_{\xv'\rightarrow \xv,\xvz'\rightarrow\xvz,t'\rightarrow t}
    \nonumber\\
  \Ja(\xv,t) & = & -\frac{e^2}{m} 
     A_{z}(\xv,t) <<c^\dagger(\xv t')c(\xv t)>>|_{t'\rightarrow t}
    \label{eq:Jlin}
\end{eqnarray}
where $\xvz$ and $\xvz'$ represent position in the lead where the 
electronic field is applied. 
Double bracket $<<>>$ includes the averaging over the 
electron and impurity.
The correlation function $Q_{\xv\xv',\xvz\xvz'}(t,t')$ is 
defined as
\begin{equation}
    Q_{\xv\xv',\xvz\xvz'}(t,t',\omega) \equiv (-i)^2 \int_{C} d\tz 
    e^{-i\omega \tz}  
    <<T_{C} c(\xv t)c^\dagger(\xvz'\tz)c(\xvz \tz)c^\dagger(\xv' t')>>
    \label{eq:Qdef}
\end{equation}
$T_{C}$ denotes a path order on contour $C$ 
in complex time plane (Fig. \ref{FIGAB_c})
and superscript $<$ denotes taking the lesser component with respect 
to $t<t'$ on the path $C$.\cite{Haug98}
Fourier transform of $Q$ is written as
\begin{equation}
    Q_{\kv\kv'}(t,t',\omega) \equiv (-i)^2 \int_{C} d\tz 
    e^{-i\omega \tz}  
    <<T_{C} c_{\kv}(t)c^\dagger_{\kv'}(\tz)c_{\kv'}(\tz)c^\dagger_{\kv}(t')>>
    \label{eq:QFT}
\end{equation}
and spatially uniform component of the current $\Jzero$ is written as  
\begin{equation}
   \Jzero(t)  = \frac{1}{V}\frac{E_{z}}{\omega}
               \frac{e^2}{2m} \sum_{\kv\kv'} 
               \frac{k_{z}k_{z}'}{m} Q^{<}_{\kv\kv'}(t,t,\omega)
    \label{eq:J0}
\end{equation}
%
We first consider a case of a simply connected geometry.
The second order contribution to $Q$ is the self-energy (SE) type
(Fig. \ref{FIGAB_se}).
(Vertex correction vanishes since 
and hence $\kv$ and $\kv'$ in eq. (\ref{eq:J0}) becomes independent 
on each other (note that $V$ does not depend on momentum transfer) 
and $Q^{<}_{\kv\kv'}$ is an even function of $\kv$ and $\kv'$.)
SE contribution, $Q^{{\rm (SE)}}_{\kv\kv'}
\equiv \delta_{\kv,\kv'}Q^{{\rm (SE)}}_{\kv}$, is written as 
\begin{eqnarray}
  Q^{{\rm (SE)}}_{\kv}(t,t',\omega)
   & =& \int_{C} d\tz e^{-i\omega \tz}  \int_{C} d t_{1} \int_{C} d t_{2} 
    \left(   G_{\kv}(t-t_{1}) \Sigma(t_{1},t_{2}) G_{\kv}(t_{2}-\tz)
    G_{\kv}(\tz-t')   \right.
    \nonumber\\
  &&  \left.  + 
  G_{\kv}(t-\tz)G_{\kv}(\tz-t_{1}) \Sigma(t_{1},t_{2}) 
    G_{\kv}(t_{2}-t') 
  \right) .
    \label{eq:QSE}
\end{eqnarray}
Here 
$\Sigma(t_{1},t_{2}) 
 \equiv i \sum_{\Qv} \chi(t_{1},t_{2})G_{\kv-\Qv}(t_{1}-t_{2})$
and 
\begin{equation}
    \chi(t_{1},t_{2})\equiv -i<T_{c} V(t_{1}) V(t_{2})>
    \label{eq:chidef}
\end{equation}
is a correlation function of TLS. 
The lesser component $Q^{<}$ (eq. (\ref{eq:QSE})) 
is calculated by use of decomposition rules such as
$[\int_{C}dt_{1} A(t-t_{1})B(t_{1}-t')]^< 
$$=\int_{-\infty}^{\infty} dt_{1} 
[A^r(t-t_{1}) B^<(t_{1}-t')+A^< (t-t_{1}) B^a(t_{1}-t')]$ 
and 
$(A(t-t_{1})B(t-t_{1}))^<$$= A(t-t_{1})^< B(t-t_{1})^<$ 
($A$ and $B$ are path-ordered correlation functions)\cite{Haug98}.
The result is (see Fig. \ref{FIGAB_seKL})
\begin{eqnarray}
  Q^{{\rm (SE)}}_{\kv}(t,t',\omega)
   & =& \sum_{\omega'} e^{-i\omega' t}  
    \left(  
    G_{\kv,\omega_{1}-\omega}^r 
    \Sigma_{\omega_{1}-\omega,\omega_{1}-\omega'}^r 
    G_{\kv,\omega_{1}-\omega'}^r G_{\kv,\omega_{1}-\omega-\omega'}^< \right.
    \nonumber\\
  &&  \left.  + 
    G_{\kv,\omega_{1}-\omega}^r 
    \Sigma_{\omega_{1}-\omega,\omega_{1}-\omega'}^r 
    G_{\kv,\omega_{1}-\omega'}^< G_{\kv,\omega_{1}-\omega-\omega'}^a 
  +     G_{\kv,\omega_{1}-\omega}^r 
    \Sigma_{\omega_{1}-\omega,\omega_{1}-\omega'}^< 
    G_{\kv,\omega_{1}-\omega'}^a G_{\kv,\omega_{1}-\omega-\omega'}^a  
    \right.\nonumber\\
  &&  \left. 
  + 
    G_{\kv,\omega_{1}-\omega}^< 
    \Sigma_{\omega_{1}-\omega,\omega_{1}-\omega'}^a 
    G_{\kv,\omega_{1}-\omega'}^a G_{\kv,\omega_{1}-\omega-\omega'}^a 
+  {\rm c.c.} 
  \right) 
\end{eqnarray}
where 
\begin{equation}
    \Sigma_{\omega_{1},\omega_{2}}^r
=i\sum_{\Qv\omega_{3}}
(G_{\kv-\Qv,\omega_{3}}^r 
\chi_{\omega_{1}-\omega_{3},\omega_{2}-\omega_{3}}^<
+G_{\kv-\Qv,\omega_{3}}^> 
\chi_{\omega_{1}-\omega_{3},\omega_{2}-\omega_{3}}^r),
\end{equation}
$\Sigma_{\omega_{1},\omega_{2}}^<
=i\sum_{\Qv\omega_{3}}$$
G_{\kv-\Qv,\omega_{3}}^< 
\chi_{\omega_{1}-\omega_{3},\omega_{2}-\omega_{3}}^<$
and 
${\rm c.c.}$ denotes conjugate processes.
Lesser and greater components of free Green functions are given as
$G^<_{\kv}(\omega)=f_{\omega}\Delta G_{\kv}(\omega)$ and 
$G^>_{\kv}(\omega)=-(1-f_{\omega})\Delta G_{\kv}(\omega)$, where
$f_{\omega}\equiv 1/(e^{\beta\omega}+1)$ is Fermi distribution 
function and 
$\Delta G_{\kv}(\omega)\equiv G^a_{\kv}(\omega)-G^r_{\kv}(\omega)$.
The expression of $Q^<$ is further simplified if we use 
\begin{equation}
\frac{k_{z}}{m} (G_{\kv}^r(\omega))^2 =\frac{\partial}{\partial k_{z}} 
G_{\kv}^r(\omega)
\end{equation}
and a partial derivative with respect to $k_{z}$. 

After some calculation SE contribution is obtained as
\begin{eqnarray}
  \lefteqn{ 
  \sum_{\kv} \frac{(k_{z})^2}{m} 
     Q^{{\rm (SE)}<}_{\kv}(t,t,\omega\rightarrow0) 
     =  -i\omega \sum_{\kv} \frac{(k_{z})^2}{m}
    \sum_{\omega'}e^{-i\omega' t} \sum_{\omega_{1}\omega_{2}} \sum_{\qv}
    }
    \nonumber\\
    &&   
   \left[ \Pi_{\kv \Qv}^{a}(\omega_{1},\omega_{2},\omega') 
    \partial_{\omega_{1}}f_{\omega_{1}} G_{\kv,\omega_{1}}^r 
    G_{\kv,\omega_{1}}^a G_{\kv,\omega_{1}-\omega'}^a
     + \Pi_{\kv \Qv}^{r}(\omega_{1},\omega_{2},\omega') 
    \partial_{\omega_{1}}f_{\omega_{1}-\omega'} 
    G_{\kv,\omega_{1}-\omega'}^a 
    G_{\kv,\omega_{1}-\omega'}^r G_{\kv,\omega_{1}}^r  \right]
    \nonumber  \\  
 & - & i \sum_{\kv} \sum_{\omega'}e^{-i\omega' t} 
     \sum_{\omega_{1}\omega_{2}} \sum_{\qv}
     \nonumber\\
    && 
    \left[ 
    \Pi_{\kv \Qv}^{a}(\omega_{1},\omega_{2},\omega')   f_{\omega_{1}}
    \Delta G_{\kv,\omega_{1}} G_{\kv,\omega_{1}-\omega'}^a
     +  \Pi_{\kv \Qv}^{r}(\omega_{1},\omega_{2},\omega') 
    f_{\omega_{1}-\omega'} 
    G_{\kv,\omega_{1}}^r  \Delta G_{\kv,\omega_{1}-\omega'}  \right.
    \nonumber  \\
    && \left.
    +\chi_{\omega_{2},\omega_{2}-\omega'}^<  
    f_{\omega_{1}-\omega_{2}} \Delta G_{\kv-\Qv,\omega_{1}-\omega_{2}}
    G_{\kv,\omega_{1}}^r  G_{\kv,\omega_{1}-\omega'}^a \right]    
    \label{eq:Jse}
\end{eqnarray}
where 
\begin{equation}
    \Pi_{\kv \Qv}^{\mu}(\omega_{1},\omega_{2},\omega')
    \equiv 
    \chi_{\omega_{2},\omega_{2}-\omega'}^< 
    G_{\kv-\Qv,\omega_{1}-\omega_{2}}^{\mu} -(1-f_{\omega_{1}-\omega_{2}}) 
    \Delta G_{\kv-\Qv,\omega_{1}-\omega_{2}}
    \chi_{\omega_{2},\omega_{2}-\omega'}^{\mu} 
\end{equation}
($\mu=a, r$) and $\sum_{\omega}\equiv\int\frac{d\omega}{2\pi}$.
The current contribution from SE, $J^{\rm (SE)}$, is defined by eq. 
(\ref{eq:J0}) with $Q$ replaced by $Q^{\rm (SE)}$.

The current $\Ja$ is similarly calculated as
\begin{eqnarray}
    \lefteqn{
      \Ja(t) =  E_{z} \frac{e^2}{m}\frac{1}{\omega} 
      \left[  \int_{C} dt_{1} \int_{C} dt_{2} 
       G_{\kv}(t-t_{1}) \Sigma(t_{1},t_{2}) G_{\kv}(t_{2}-t')  
       \right]^<  
       }
    \nonumber \\ 
     && = iE_{z} \frac{e^2}{m}\frac{1}{\omega} 
     \sum_{\kv} \sum_{\omega'}e^{-i\omega' t} 
     \sum_{\omega_{1}\omega_{2}} \sum_{\Qv}
     \nonumber\\
    && 
    \left[ 
    \Pi_{\kv \Qv}^{a}(\omega_{1},\omega_{2},\omega') f_{\omega_{1}}
    \Delta G_{\kv,\omega_{1}} G_{\kv,\omega_{1}-\omega'}^a
     +  \Pi_{\kv \Qv}^{r}(\omega_{1},\omega_{2},\omega')
    f_{\omega_{1}-\omega'} 
    G_{\kv,\omega_{1}}^r  \Delta G_{\kv,\omega_{1}-\omega'}  \right.
    \nonumber  \\
    && \left.
    +\chi_{\omega_{2},\omega_{2}-\omega'}^<  
    f_{\omega_{1}-\omega_{2}} \Delta G_{\kv-\Qv,\omega_{1}-\omega_{2}}
    G_{\kv,\omega_{1}}^r  G_{\kv,\omega_{1}-\omega'}^a \right]    
       \label{eq:Ja} 
\end{eqnarray}
It is seen that this contribution cancels the second part in 
eq. (\ref{eq:Jse}). Hence the total current is obtained as
\begin{eqnarray}
  \lefteqn{ J(t)= J_{0}+J^{\rm (SE)}(t)+\Ja(t)  }
  \nonumber\\
  &=&
  J_{0} -i \frac{1}{V} E_{z}
    \left(\frac{e}{m}\right)^2 \sum_{\kv} (k_{z})^2
    \sum_{\omega'}e^{-i\omega' t} \sum_{\omega_{1}\omega_{2}} \sum_{\Qv}
    \nonumber\\
    &&   
   \left[  \Pi_{\kv \Qv}^{a}(\omega_{1},\omega_{2},\omega')
    \partial_{\omega_{1}}f_{\omega_{1}} G_{\kv,\omega_{1}}^r 
    G_{\kv,\omega_{1}}^a G_{\kv,\omega_{1}-\omega'}^a
     +   \Pi_{\kv \Qv}^{r}(\omega_{1},\omega_{2},\omega')
    \partial_{\omega_{1}}f_{\omega_{1}-\omega'} 
    G_{\kv,\omega_{1}-\omega'}^a 
    G_{\kv,\omega_{1}-\omega'}^r G_{\kv,\omega_{1}}^r  \right]
\label{eq:Jresult0}
\end{eqnarray}
Here $J_{0}\equiv E_{z}\sigma_{0}$ is current without TLS, 
$\sigma_{0}\equiv\frac{e^2}{3}\left( \frac{k_{F}}{m} 
\right)^2\frac{\DOS}{V}\tau$, $\DOS\equiv V(mk_{F}/2\pi^2)$ is the 
density of states. 

Using $\partial_{\omega_{1}}f_{\omega_{1}}\simeq -\delta(\omega_{1})$ 
and taking summations over $\kv$ and $\Qv$, we obtain
\begin{equation}
   J(t)=J_{0}- 2\pi J_{0} \DOS\tau
    \sum_{\omega'\omega_{2}}\frac{e^{-i\omega' t}}{1-i\tau\omega'} 
   i\left[ \chi_{\omega_{2},\omega_{2}-\omega'}^< 
     -f_{\omega_{2}} \chi_{\omega_{2},\omega_{2}-\omega'}^a
     +f_{\omega_{2}-\omega'} \chi_{\omega_{2},\omega_{2}-\omega'}^r
    \right]
    \label{eq:Jresult1}
\end{equation}
\subsection{Aharonov-Bohm current}
\begin{figure}[bthp]
\begin{center}
\epsfxsize=10cm              
\epsfbox{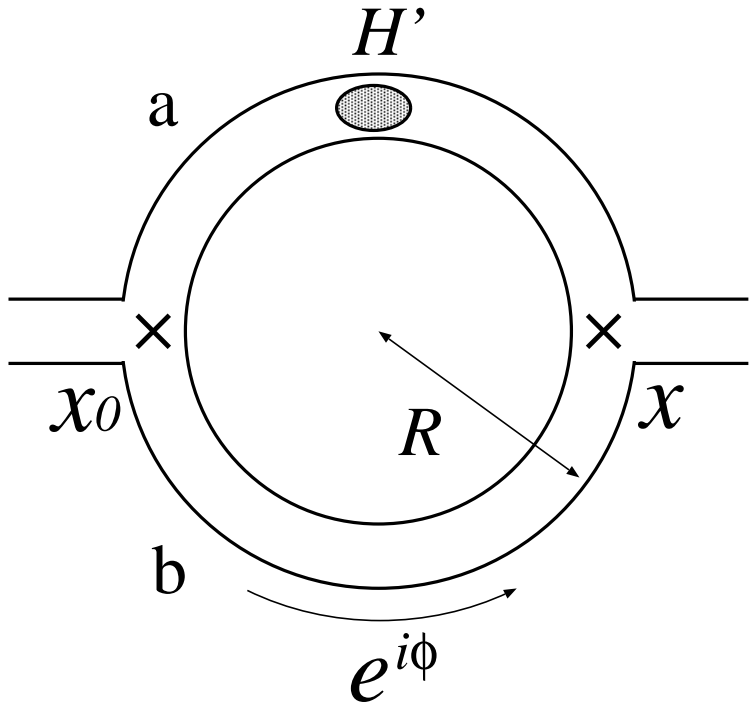}
\end{center}
\caption{Ring we consider. 
TLS affects only the upper arm (\arma) and the phase due to magnetic 
flux ($\phase$) is attached only on the lower arm, \armb.
\label{FIGAB_ring}}
\end{figure}

We next consider the case of a ring with a magnetic flux, shown in 
Fig. \ref{FIGAB_ring}.
For simplicity the perturbation due to TLS ($H'$) is treated as to 
exist only on the upper arm (arm {\arma}) and 
the phase $\phase\equiv 2\pi\Phi/\Phi_{0}$ ($\Phi_{0}\equiv h/2e$ 
being flux quantum) due to the flux ($\Phi$) 
affects only the lower arm, {\armb}. 
We consider the case the ring is slowly varying and the system is ballistic, 
$L\lesssim \ell$.
The current through the ring is given by the same expression as
eqs.  (\ref{eq:Jlin})-(\ref{eq:J0})  but green functions need to be replaced 
by those in the ring geometry.
%
Green function connecting $\xv$ and $\xvz$ at the right and left end 
of the ring, respectively, is approximated as
\begin{eqnarray}
    G_{\rm ring}(\xv-\xvz) & \simeq & G_{\arma}(\xv-\xvz)+G_{\armb}(\xv-\xvz)
    \nonumber \\
     & = & [G(\xv-\xvz)+(G\Sigma G)(\xv-\xvz)]+e^{i\phase}G(\xv-\xvz)
    \label{eq:Gfring1}
\end{eqnarray}
where the first term is the Green function though the arm {\arma} 
($G_{\arma}\equiv G+G\Sigma G$) 
and $G_{\armb}(\xv-\xvz)\equiv e^{i\phase}G(\xv-\xvz)$ 
represents propagation through arm {\armb}.
In eq. (\ref{eq:Gfring1}) contributions from the multiple circulation 
through the ring is neglected.
The Green function in the opposite direction from $\xv$ to $\xvz$ is 
\begin{equation}
    G(\xvz-\xv)  =  G_{\arma}(\xvz-\xv)+G_{\bar{\armb}}(\xvz-\xv)
    \label{eq:Gfring2}
\end{equation}
where $G_{\bar{\armb}}(\xvz-\xv)\equiv e^{-i\phase}G(\xvz-\xv)$ carries the 
opposite phase as $G_{\armb}$.
The current through the ring is calculated from eq. (\ref{eq:Jlin})  as
\begin{equation}
     J_{\rm ring}(t) \equiv  J_{\arma}+J_{\armb}+J_{\arma\armb}
    \label{eq:Jring}
\end{equation}
where $J_{\arma}$ and $J_{\armb}$ are ($\alpha=\arma,\armb$)
\begin{eqnarray}
   J_{\alpha} & = & \frac{1}{V}\frac{E_{z}}{\omega}
         \left(\frac{e}{m}\right)^2 \sum_{\kv}  {k_{z}^2} 
      \sum_{\omega'}e^{-i\omega' t}\sum_{\omega_{1}\omega_{2}} 
    \nonumber \\
     &  & \left[ 
     {G_{\alpha}}_{\kv}^{r} (\omega_{1}+\omega',\omega_{2}+\omega)
     {G_{\bar{\alpha}}}_{\kv}^{<} (\omega_{2},\omega_{1})	
    +{G_{\alpha}}_{\kv}^{<}(\omega_{2},\omega_{1})
     {G_{\bar{\alpha}}}_{\kv}^{a} (\omega_{1}-\omega,\omega_{2}-\omega') 
         \right]
	+{\Ja}_{\alpha}
     \label{eq:Jring0}
\end{eqnarray}
${\Ja}_{\alpha}$ being contribution from ${\Ja}$ on arm $\alpha$ and 
$G_{\bar{\arma}}\equiv G_{\arma}$.
Current through arm {\arma}, $J_{\arma}$, is equal at 
(\ref{eq:Jresult1}) and $J_{\armb}=J_{0}$ since $H'=0$ on arm \armb. 
Fourier transform of Green functions in eq. (\ref{eq:Jring0}) are defined as
${G_{\alpha}}_{\kv}^{\mu} (\omega_{1},\omega_{2}) \equiv
\int_{-\infty}^{\infty} dt_{1} \int_{-\infty}^{\infty} dt_{2}
e^{-i\omega_{1}t_{1}} e^{i\omega_{2}t_{2}}  {G_{\alpha}}_{\kv}^{\mu} 
(t_{1},t_{2})$. 
($\omega_{1}$ is not necessarily equal to $\omega_{2}$ since 
$G_{\alpha}$ includes the self-energy due to TLS, which is not 
energy-conserving). 
In (\ref{eq:Jring}), the interference effect is included in 
$J_{\arma\armb}$, 
which reads 
\begin{eqnarray}
    J_{\arma\armb} & = & \frac{1}{V}\frac{E_{z}}{\omega}
         \left(\frac{e}{m}\right)^2 \sum_{\kv}  {k_{z}^2} 
      \sum_{\omega'}e^{-i\omega' t}\sum_{\omega_{1}} 
    \nonumber \\
     &  & \left[ 
     {G_{\arma}}_{\kv}^{r}
           (\omega_{1}-\omega,\omega_{1}-\omega')
     {G_{\bar{\armb}}}_{\kv}^{<}
           (\omega_{1})
     +
    {G_{\arma}}_{\kv}^{<}
           (\omega_{1}-\omega,\omega_{1}-\omega')
     {G_{\bar{\armb}}}_{\kv}^{a}
           (\omega_{1}-\omega-\omega') 
      \right.   \nonumber\\
   & &  \left.   
    + 
	{G_{{\armb}}}_{\kv}^{r}
           (\omega_{1}) 
	{G_{\arma}}_{\kv}^{<}
           (\omega_{1}-\omega,\omega_{1}-\omega') 
     + {G_{{\armb}}}_{\kv}^{<}
           (\omega_{1}) 
       {G_{\arma}}_{\kv}^{a}
           (\omega_{1}-\omega,\omega_{1}-\omega') 
     \right]
     +\Ja_{\arma\armb}
     \nonumber\\
    &\equiv& J_{\arma\armb}^{\rm (0)}+J_{\arma\armb}^{\rm (2)}
     \label{eq:JringAB}
\end{eqnarray}
Here
$
    J_{\arma\armb}^{\rm (0)}  =  2J_{0}\cos\phase
$
and (by use of (\ref{eq:Gfring1}) and (\ref{eq:Gfring2}))
\begin{equation}
    J_{\arma\armb}^{\rm (2)}  =   \frac{1}{V}\frac{E_{z}}{\omega}
         \left(\frac{e}{m}\right)^2 \sum_{\kv}  {k_{z}^2} 
      \sum_{\omega'}e^{-i\omega t}
      \left[ e^{i\phase} Q_{\kv}^{+<}(\omega,\omega')+
             e^{-i\phase} Q_{\kv}^{-<}(\omega,\omega') \right]
	  +{\Ja}_{\arma\armb}^{\rm (2)}
    \label{eq:JAB}
\end{equation}
where
\begin{eqnarray}
    Q_{\kv}^{+<}(\omega,\omega') &\equiv& \sum_{\omega_{1}}
    \left[ G_{\kv,\omega_{1}}^{r}
    (G\Sigma G)^<_{\omega_{1}-\omega,\omega_{1}-\omega'} 
   + G_{\kv,\omega_{1}}^{<}
    (G\Sigma G)^a_{\omega_{1}-\omega,\omega_{1}-\omega'} \right]
    \nonumber\\ 
    Q_{\kv}^{-<}(\omega,\omega') &\equiv& \sum_{\omega_{1}}
    \left[  (G\Sigma G)^r_{\omega_{1}-\omega,\omega_{1}-\omega'} 
    G_{\kv,\omega_{1}-\omega-\omega'}^{<}
   + (G\Sigma G)^<_{\omega_{1}-\omega,\omega_{1}-\omega'} 
     G_{\kv,\omega_{1}-\omega-\omega'}^{a} \right]
    \label{eq:Qpm}
\end{eqnarray}
These are calculated similarly to the derivation of eq. (\ref{eq:Jse}) as
\begin{eqnarray}
    Q_{\kv}^{+<}(\omega\rightarrow0,\omega') 
    & = & -i\omega\sum_{\omega_{1}\omega_{2}}
    \partial_{\omega_{1}}f_{\omega_{1}} G_{\kv,\omega_{1}}^{r}
     G_{\kv,\omega_{1}}^{a} G_{\kv,\omega_{1}-\omega'}^{a}
   \Pi_{\kv \Qv}^{a}(\omega_{1},\omega_{2},\omega')
   +{Q^+}'
   \nonumber\\
    Q_{\kv}^{-<}(\omega\rightarrow0,\omega')     
    & = & -i\omega\sum_{\omega_{1}\omega_{2}}
    \partial_{\omega_{1}}f_{\omega_{1}-\omega'} G_{\kv,\omega_{1}}^{r}
     G_{\kv,\omega_{1}-\omega'}^{r} G_{\kv,\omega_{1}-\omega'}^{a}
     \Pi_{\kv \Qv}^{r}(\omega_{1},\omega_{2},\omega')
     +{Q^-}'
        \label{eq:Qpm2}
\end{eqnarray}
where ${Q^\pm}'$ are terms which cancel with ${\Ja}_{\arma\armb}^{\rm (2)}$.
Hence $J_{\arma\armb}^{\rm (2)}$ is obtained as
\begin{eqnarray}
    J_{\arma\armb}^{\rm (2)} & = &  -\frac{2\pi^2}{3} 
     \frac{1}{V} {E_{z}}
         \left(\frac{e}{m}\right)^2 (\DOS k_{F} \tau)^2 
      \sum_{\omega'}\frac{e^{-i\omega' t}}{1-i\omega'\tau}  
      \sum_{\omega_{2}}
      \nonumber\\
     &&
     \times i \frac{1}{2} \left[ e^{i\phase} 
       (\chi^<_{\omega_{2},\omega_{2}-\omega'} 
          -2f_{\omega_{2}} \chi^a_{\omega_{2},\omega_{2}-\omega'} )
            +e^{-i\phase} 
	(\chi^<_{\omega_{2},\omega_{2}-\omega'} 
          +2f_{\omega_{2}-\omega'} \chi^r_{\omega_{2},\omega_{2}-\omega'} ) 
	  \right]
	  \label{eq:JAB2}
\end{eqnarray}
From eqs. (\ref{eq:Jring0}) and (\ref{eq:JAB2}), we obtain the total 
current through the ring as
\begin{eqnarray}
    \lefteqn{ 
    J_{\rm ring} (t) =   2(1+\cos\phase) J_{0}  
      -2\pi J_{0} \DOS \tau
      \sum_{\omega'}\frac{e^{-i\omega' t}}{1-i\omega'\tau}  
      \sum_{\omega_{2}}
      } \nonumber\\
  && 
     \times  \frac{i}{2} \left[ (1+e^{i\phase} )
       (\chi^<_{\omega_{2},\omega_{2}-\omega'} 
          -2f_{\omega_{2}} \chi^a_{\omega_{2},\omega_{2}-\omega'} )
            + (1+e^{-i\phase} )
	(\chi^<_{\omega_{2},\omega_{2}-\omega'} 
          +2f_{\omega_{2}-\omega'} \chi^r_{\omega_{2},\omega_{2}-\omega'} ) 
	  \right]
	  \label{eq:JAB3}
\end{eqnarray}

\section{Correlation functions of Two-Level-System}
The Hamiltonian of TLS we consider is  
\begin{equation}
    H_{\rm TLS}=\frac{\Omega}{2} \sigma_{z}
    \label{eq:HTLS}
\end{equation}
where the two levels are represented by Pauli matrix $\sigma_{z}$. 
The interaction $H'$ is switched on at $t=0$ till $t=\tm$ 
($\tm$ is later set equal to the time of measurement), and is written 
as
\begin{equation}
    V(t)=(u \sigma_{z} + v \sigma_{x}) \theta(t)\theta(\tm-t)
    \label{eq:Vtdef}
\end{equation}
where $u,v$ are coupling constants and $\theta(t)$ is a step 
function.
Here we consider the case TLS is initially at $|\sigma_{z}=m>$ 
($m=\pm1$) at $t=0$.
The correlation functions are given as
\begin{eqnarray}
    \chi^< (t_{1},t_{2}) & = & -i<m| V(t_{2}) V(t_{1})|m>
    \nonumber  \\
     & = & -i(u^2+v^2 e^{-im\Omega(t_{1}-t_{2})}) 
     \theta(t_{1})\theta(t_{2}) \theta(\tm-t_{1})\theta(\tm-t_{2})
    \nonumber  \\
     \chi^> (t_{1},t_{2})  & = & -i(u^2+v^2 
     e^{im\Omega(t_{1}-t_{2})}) \theta(t_{1})\theta(t_{2}) 
     \theta(\tm-t_{1})\theta(\tm-t_{2}) 
     \nonumber\\
     \chi^r(t_{1},t_{2})&=&\theta(t-t')(\chi^>-\chi^<)(t_{1},t_{2}),
     \nonumber\\
     \chi^a(t_{1},t_{2})&=& -\theta(t'-t)(\chi^>-\chi^<)(t_{1},t_{2})
     \label{eq:chis} 
\end{eqnarray}
Fourier transform is defined as ($\mu=<,>,r,a$)
\begin{equation}
    \chi^\mu_{\omega,\omega'}\equiv \int_{-\infty}^\infty 
    dt_{1}\int_{-\infty}^\infty dt_{2} 
    e^{i\omega t_{1}}e^{-i\omega' t_{2}} \chi^\mu(t_{1},t_{2})
    \label{eq:chift}
\end{equation}
These are calculated as (note that these Fourier transforms contains 
time $\tm$)
\begin{eqnarray}
    \chi_{\omega_{2},\omega_{2}-\omega'}^< & = & -i[
   u^2 \Gamma_{\omega_{2}} \Gamma_{\omega'-\omega_{2}} 
  +v^2 
  \Gamma_{\omega_{2}-m\Omega}\Gamma_{\omega'-\omega_{2}+m\Omega} ] 
    \nonumber \\
     \chi_{\omega_{2},\omega_{2}-\omega'}^a & = & 
      v^2 \sum_{\pm} 
     \frac{\pm}{\omega_{2}\pm m\Omega}
     (\Gamma_{\omega'}-\Gamma_{\omega'-\omega_{2}\mp m\Omega})
    \nonumber  \\
     \chi_{\omega_{2},\omega_{2}-\omega'}^r & = & 
     v^2 \sum_{\pm} 
     \frac{\pm}{\omega_{2}-\omega'\pm m\Omega}
     (\Gamma_{\omega'}-\Gamma_{\omega_{2}\pm m\Omega})
     \label{eq:chisft}
\end{eqnarray}
where 
\begin{equation}
    \Gamma_{\omega}(\tm)\equiv \int_{0}^{\tm} dt' e^{i\omega t'} 
    = \frac{e^{i\omega \tm}-1}{i(\omega+i0)}
\end{equation}

\section{Response of Aharonov-Bohm current to TLS}

\begin{figure}[bthp]
\begin{center}
\epsfxsize=10cm              
\epsfbox{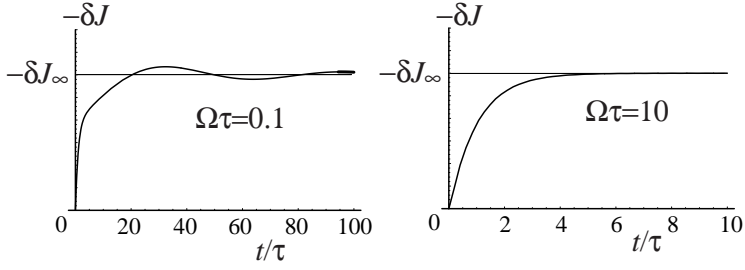}
\caption{
Behavior of the current, eq. (\ref{eq:Jresult2}) for the two 
cases $\tilM=0.1$ and $\tilM=10$. $u$ and $v$ are chosen as 1.
\label{FIGAB_Jt}}
\end{center}
\end{figure}

The expression of (\ref{eq:Jresult1}) is estimated by use of 
(\ref{eq:chisft}) with $\tm\rightarrow t$ as
\begin{eqnarray}
   \lefteqn{ J(t) = J_{0} } 
       \nonumber  \\
     &  &
     -2\pi J_{0} \DOS \tau
     \left[ 
      \left\{ u^2+v^2\left(1-\frac{2}{\pi}\tan^{-1}\tilM \right) \right\} 
       (1-e^{-\tilt})   \right.
 \left. +\frac{2}{\pi}v^2 
     \int_{0}^{\tilM}\frac{dx}{1+x^2}
     \left(1-\cos x\tilt +\frac{\sin x\tilt}{x} \right) \right]
    \label{eq:Jresult2}
\end{eqnarray}
where $\tilM\equiv m\Omega\tau$, $\tilt\equiv t/\tau$. 
In the case of low frequency, $|\tilM|\ll 1$,
integration over $x$ is carried out to be
\begin{equation}
      J(t) = J_{0}\left[1
      -2\pi\DOS \tau 
      \left\{ \left(u^2+v^2\right) (1-e^{-\tilt}) 
     +v^2\frac{2}{\pi}
      \left( {\rm Si}(\tilM \tilt)-\frac{\sin\tilM \tilt}{\tilt} 
      +\tilM e^{-\tilt} \right) \right\} \right]
        \;\;\;\;\; (|\tilM|\ll 1)
    \label{eq:JsmallM}
\end{equation}
where ${\rm Si}(x)\equiv \int_{0}^{x} (dy/y)\sin y$.
After TLS ($H'$) is switched on, the current relaxes to a new equilibrium 
value ($J_{0}\left[1-2\pi\DOS \tau \left(u^2+2v^2\right) \right]
\equiv J_{0}+\delta J_{\infty}$) 
 in the time scale of $\Omega^{-1}$ (Fig. \ref{FIGAB_Jt}).
In the opposite case, $|\tilM|\gg 1$, the scale becomes $\tau$; 
\begin{equation}
      J(t) = J_{0}\left[ 1- 2\pi\DOS\tau
     \left(u^2+2v^2\right) (1-e^{-\tilt})  \right] 
        \;\;\;\;\; (|\tilM|\gg 1)
    \label{eq:JlargeM}
\end{equation}

The result for the ring, eq. (\ref{eq:JAB3}), is similarly calculated 
as
\begin{eqnarray}
    \lefteqn{ 
    J_{\rm ring} (t) =  2 (1+\cos\phase) J_{0} 
    -(1+\cos\phase) J_{0} 2\pi\DOS\tau}
   \nonumber\\
  && 
 \times
     \left[  \left\{ 
      u^2+v^2\left(1-\frac{2}{\pi}\tan^{-1}\tilM \right)  \right\}
     (1-e^{-\tilt})   \right.
     \left.
     +\frac{2}{\pi}v^2 
     \int_{0}^{\tilM}\frac{dx}{1+x^2}
     \left(1-\cos x\tilt +\frac{\sin x\tilt}{x} \right) \right]
	  \label{eq:JAB4}
\end{eqnarray}
It is seen that the amplitude of AB oscillation is reduced by TLS but 
the reduction is due to the  reflection of 
the electron by TLS in the same way as in a wire (eq. 
(\ref{eq:Jresult2})).
Namely the interference is not affected by the 
TLS in ballistic transport. 
This is also seen from the effect surviving even in the limit of 
$\Omega\rightarrow0$ and $\rightarrow\infty$ (compare with the AAS 
current, Eq. (\ref{Fdef})).

The phase of the oscillation ($\cos\phase$) is not modified,
similarly to the 
equilibrium case, in which case $\sin\phase$ term is forbidden since it 
violates the time-reversal symmetry\cite{Yacoby96}.


The behavior at $t\sim0$ of the current (\ref{eq:JAB4}) is given as 
\begin{equation}
    J_{\rm ring}(t) \simeq  (1+\cos\phase) J_{0}
    [2-2\pi\DOS(u^2+v^2) t] \simeq  (1+\cos\phase) 2J_{0} e^{-\Gamma 
    t/2}
    \label{eq:Jtzero}
\end{equation}
where $\Gamma\equiv 2\pi\DOS(u^2+v^2)$ and a factor of $1/2$ is to 
account for the TLS applied only on one of the two arms. 
This decay rate $\Gamma$ is nothing but the rate obtained by Fermi's 
golden rule. In fact transition probability of the electron from 
momentum $\kv$ to $\kv'$ is given by 
\begin{eqnarray}
    |A_{\kv'm'\kv m}(t)|^2 
     & =& \left| -i\int_{0}^t dt_{1} <\kv'm'|H'(t_{1})|\kv m>  \right|^2
     \nonumber\\
     &=&u^2 \delta_{m'm} \left( 
       \frac{\sin[(\epsilon_{\kv'}-\epsilon_{\kv})t/2]} 
             {(\epsilon_{\kv'}-\epsilon_{\kv})/2} \right)^2
      + v^2 \delta_{m',-m} \left( 
       \frac{\sin[(\epsilon_{\kv'}-\epsilon_{\kv}-m\Omega)t/2]} 
             {(\epsilon_{\kv'}-\epsilon_{\kv}-m\Omega)/2} \right)^2
    \label{eq:Asq}
\end{eqnarray}
where $m$ and $m'(=\pm)$ are the initial and final state of TLS.
By use of $\left( \frac{\sin[\epsilon t/2]} {\epsilon/2} \right)^2 
             \rightarrow 2\pi\delta(\epsilon)t$ for $t\rightarrow\infty$, the 
rate estimated by golden rule is seen to be equal to $\Gamma$.

This rate $\Gamma$ is also evaluated from the overlap of the state at $t$ and 
$t=0$,
\begin{eqnarray}
    <0|Te^{-i\int_{0}^t dt_{1}H'(t_{1})}|0> 
    &\simeq& 1-\frac{i}{2} \int_{0}^t dt_{2}\int_{0}^t dt_{2} 
    \theta(t_{1}-t_{2})(\chi^>(t_{1},t_{2})+\chi^<(t_{2},t_{1})) 
    \nonumber\\
    && \times
    \sum_{\kv\kv'} G_{\kv'}^<(t_{2}-t_{1}) G_{\kv}^>(t_{1}-t_{2})
    \label{eq:overlap}
\end{eqnarray}
which results in $\simeq e^{-\Gamma t}$ for $\Gamma t \ll 1$.

In the case of electron-electron interaction, decay rate of the overlap 
integral was shown to be equivalent to dephasing time\cite{Stern90,Imry97}.
In the present case of ballistic transport, the decay of the 
amplitude of AB oscillation (eq. (\ref{eq:JAB4})) as well as the 
overlap integral are not related to dephasing, 
but are due simply to the scattering into other states.
What is crucial here is lack of randomness need to put uncertain phase 
on wave function. 
Dephasing is taken into account when effect of cooperon is 
considered in the presence of random disorder (section \ref{SECAAS}).

\section{Effect of oscillating external field \label{SECAC}}
\begin{figure}[bthp]
\begin{center}
\epsfxsize=10cm              
\epsfbox{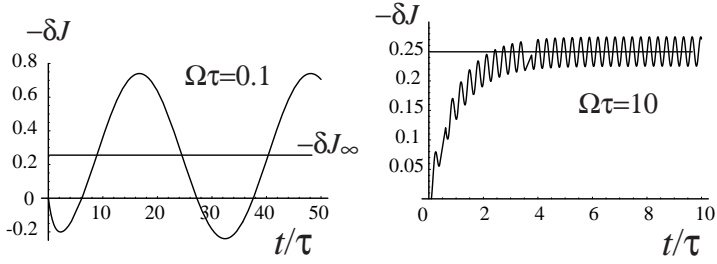}
\caption{
Behavior of $|\delta J|\equiv |J(t)-J_{0}|$ for the oscillating 
external field (eq. (\ref{eq:JAC})), 
plotted in unit of $2\pi J_0 \DOS \tau v^2$.
At $t\rightarrow\infty$ the oscillation is around $|\delta 
J|=1/4\times 2\pi J_0 \DOS \tau v^2$.
\label{FIGAB_AC}}
\end{center}
\end{figure}

Our ballistic results (\ref{eq:Jresult1}) (\ref{eq:JAB3}) are general 
and can be applied to other perturbation sources.
We here consider the current  (\ref{eq:Jresult1}) 
with an oscillating external field, $V(t)=v \sin \omega t $. 
In this case $\chi^a=\chi^r=0$ and
\begin{equation}
    \chi^<_{\omega_{2},\omega_{2}-\omega'}=\frac{i}{4}v^2 \sum_{\pm}
    \left[ \Gamma_{\omega_{2}\pm \Omega} 
     \left( \Gamma_{-\omega_{2}+\omega'\pm\Omega}- 
     \Gamma_{-\omega_{2}+\omega'\mp\Omega} \right)\right]
    \label{eq:chiac}
\end{equation}
The current (\ref{eq:Jresult1}) is obtained as
\begin{eqnarray}
    J(t) & = & J_{0} -2\pi J_0 \DOS \tau 
    \frac{v^2}{4}\frac{1}{1+4\tilM^2} 
       \nonumber \\
     &  & \times\left[2(1-\cos2\Omega t 
     -2\tilM \sin 2\Omega t)  -(1-4\tilM^2)(1-e^{-\tilt}) \right] 
    \label{eq:JAC}
\end{eqnarray}
As seen in Fig. \ref{FIGAB_AC}, the current oscillates around new 
equilibrium value ($J_{\infty}=J_{0} -2\pi J_0 \DOS \tau {v^2}/{4}$)
if the 
external field is slowly varying ($\tilM \ll 1$) but
oscillation is not dominant if the perturbation is too fast for the 
electron to accommodate ($\tilM \gg 1$). 

This result has possibility of various applications.
We mention here a case of ballistic transport through a nano-scale 
metallic magnetic contacts. 
In magnetic contacts large magnetoresistance is observed due to a 
strong scattering by a domain wall trapped in the contact 
region\cite{Garcia99,TZMG99}. 
Recently non-linear $I$-$V$ characteristic was observed in 
half-metallic oxide contacts, which is 
argued to be due to deformation of the wall\cite{Versluijs01}.
In these small contacts, application of a small oscillating magnetic field 
might drive slow oscillation of the wall position and shape. 
This causes a time varying scattering potential of the electron and 
hence would be detectable by measuring time-resolved current through 
the contact.
Current measurement may be useful to observe mesoscopic dynamics.
\section{Response of Altshuler-Aronov-Spivak oscillation 
\label{SECAAS}}
In this section we study the effect of switching of TLS on 
Altshuler-Aronov-Spivak (AAS) oscillation\cite{Altshuler81}. 
This oscillation is due to interference 
of a particle-particle propagator (Cooperon) induced by successive 
elastic scattering.
The oscillation is $\cos(2\phi)$, 
reflecting a charge of $2e$ Cooperon carries.
AAS contribution is calculated from eq. (\ref{eq:Jlin}) with Cooperon 
taken into account. 
In the absence of TLS, the Cooperon contribution to the current is 
calculated as\cite{Aronov87}
\begin{equation}
    J_{AAS}^{\rm (0)} =\frac{E_{z}}{V} \left( \frac{e}{m} \right)^2  
     \nimp\vi^2
    \sum_{\kv} k_{z}(-k_{z})  G_{\kv}^r G_{-\kv}^r G_{\kv}^a G_{-\kv}^a 
    C(0)
    \label{eq:JAAS0}
\end{equation}
where
\begin{eqnarray}
    C(0) & \equiv & \sum_{\pv} 
      \sum_{n=0}^{\infty}
      (\nimp\vi^2 \sum_{\kv}G_{\kv}^{r} G_{\pv-\kv}^{a})^n
    \nonumber \\
     & \simeq & \sum_{\pv} 
     \frac{1}{ (Dp^2+1/\tau_{\varphi}) \tau}
    \label{eq:Cdef}
\end{eqnarray}
is Cooperon. $\nimp$ and $\vi$ are density and strength of 
impurity scattering, respectively, which are related to $\tau$ as
$1/\tau=2\pi \nimp\vi^2 \DOS$. 
We have added phenomenologically an inelastic lifetime, 
$\tau_{\varphi}$\cite{Bergmann84}, which is assumed to arise from 
other mechanism than TLS.
For $L\gtrsim \ell_{\varphi}$ 
($\ell_{\varphi}\equiv\sqrt{D\tau_{\varphi}}$ is the inelastic mean 
free path (dephasing length)), $C(0)$ is calculated 
as (we assume that the width of the ring is smaller than inelastic 
mean free path ($L_{\perp}\lesssim\ell_{\varphi}$) and carry out 
summation over $p$ as in one-dimension)
\begin{equation}
   C(0)\simeq \frac{3L\ell_{\varphi}}{8\pi^2\ell^2}
    \left( 1+ 2e^{-L/\ell_{\varphi}} \cos 2\phase \right)
    \label{eq:C12}
\end{equation}
(Higher order contributions, $\propto  e^{-nL/\ell_{\varphi}}$, 
$n\geq2$ are neglected.)
The AAS current in the absence of TLS is thus  
\begin{equation}
    J_{AAS}^{\rm (0)} =-E_{z}\sigma_{0}\frac{3}{2\pi k_{F}^2 ab} 
    \frac{\ell_{\varphi}}{\ell}  e^{-L/\ell_{\varphi}} \cos 2\phase
    \label{eq:JAAS1}
\end{equation}
$a$ and $b$ being width and thickness of the ring, respectively.

\begin{figure}[bthp]
\begin{center}
\epsfxsize=10cm              
\epsfbox{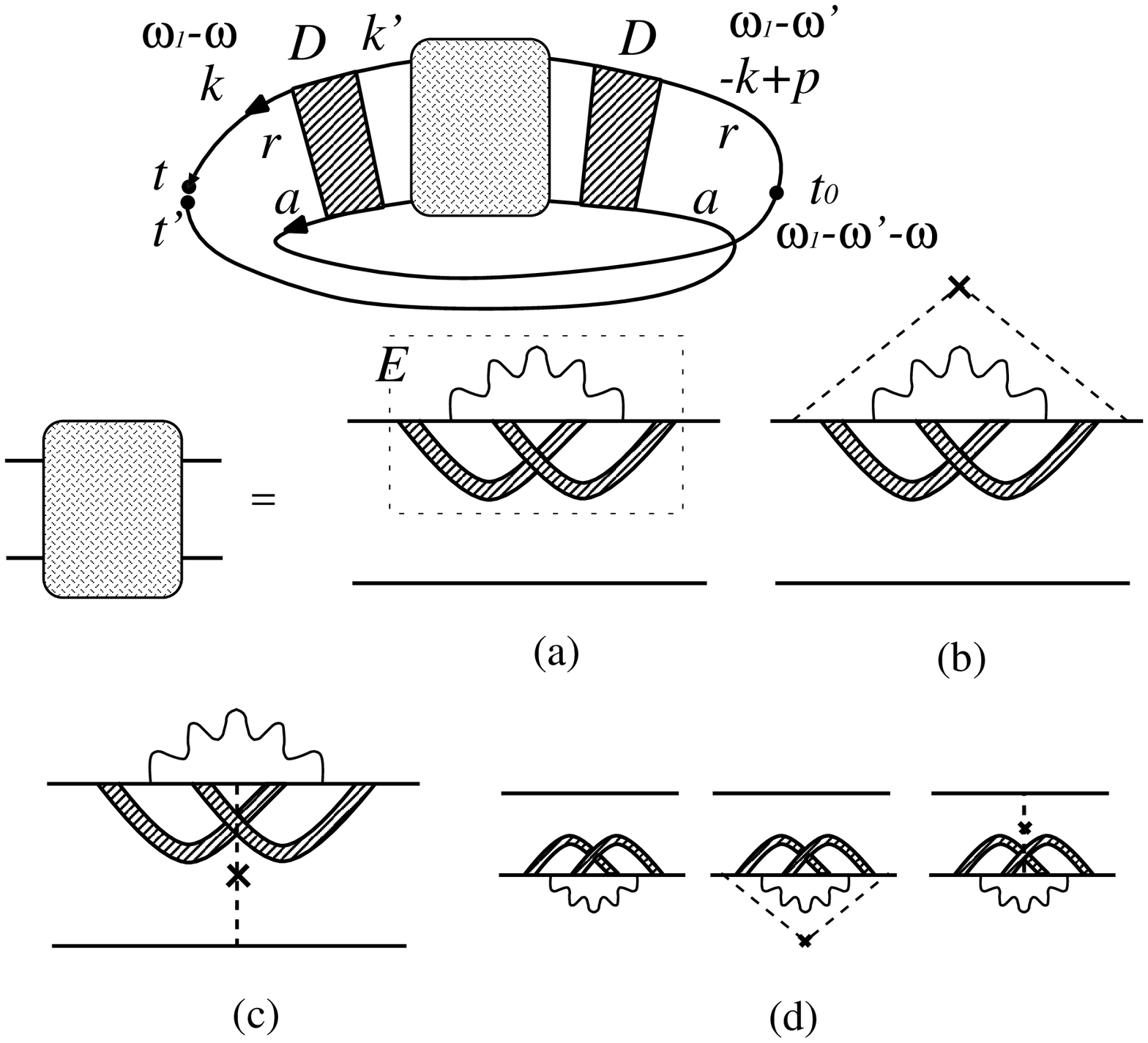}
\end{center}
\caption{Corrections by TLS to AAS oscillation. Shaded thick line 
denotes Cooperon. Scattering by normal impurity is indicated by a 
dotted line.
\label{FIGAB_AAS}}
\end{figure}

\begin{figure}[bthp]
\begin{center}
\epsfxsize=10cm              
\epsfbox{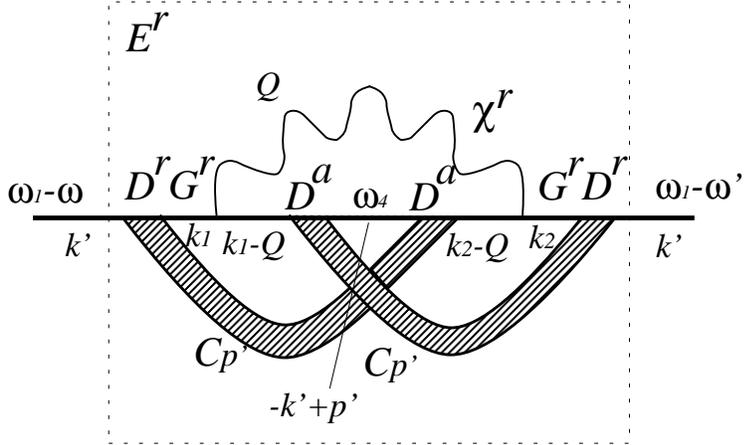}
\end{center}
\caption{Diagrammatic representation of $E^r$.
\label{FIGAB_AASKL}}
\end{figure}

Now we calculate the effect by TLS. This is done by 
considering a correction to Cooperon. 
Most important processes are shown in Fig. \ref{FIGAB_AAS}(a-c). 
Process (a) is calculated as
\begin{eqnarray}
    Q_{\kv}^{{\rm (a)}<} (t,t,\omega) & = & 
         (\nimp\vi^2)^2 \sum_{\omega' \omega_{1}} 
      e^{-i\omega' t} 
       \sum_{nn'=0}^{\infty} \sum_{\kv' \kv_{i} 
      \kv_{i}'} \sum_{\pv} 
      \nonumber\\
    && \times 
      \left[
      G_{\kv,\omega_{1}-\omega} D^{(n)}_{ \{\kv_{i}\},\omega_{1}-\omega } 
      G_{\kv',\omega_{1}-\omega} 
      E_{\omega_{1}-\omega,\omega_{1}-\omega'} 
      G_{\kv',\omega_{1}-\omega'} D^{(n')}_{ \{\kv_{i}'\},\omega_{1}-\omega'} 
      G_{-\kv+\pv,\omega_{1}-\omega'} 
      \right. \nonumber\\
    && \times \left.
      G_{-\kv+\pv,\omega_{1}-\omega-\omega'}
      D^{(n)}_{ \{ -\kv_{i}+\pv \},\omega_{1}-\omega-\omega'} 
      G_{-\kv'+\pv,\omega_{1}-\omega-\omega'} 
      D^{(n')}_{ \{ -\kv_{i}'+\pv \},\omega_{1}-\omega-\omega'} 
      G_{\kv,\omega_{1}-\omega-\omega'} \right]^{<}
     \label{eq:QcoopSE}
\end{eqnarray}
where 
\begin{equation}
    D^{(n)}_{ \{\kv_{i}\}, \omega_{1}}\equiv \Pi_{i=1}^{n} 
    (\sqrt{\nimp\vi^2} G_{\kv_{i},\omega_{1}})
    \label{eq:Ddef}
\end{equation}
is Green functions connected by successive impurity scattering,
\begin{equation}
    E(t,t')\equiv  (\nimp\vi^2)^2
     \sum_{nn'=0}^{\infty}  \int_{C}dt_{1}\int_{C} dt_{2}
    [D^{(n)}G](t-t_{1})  i \chi(t_{1},t_{2}) F^{nn'}(t_{1}-t_{2})
    [GD^{(n')}](t_{2}-t')
    \label{eq:Edef}
\end{equation}
and $F^{nn'}(t_{1}-t_{2})\equiv [GD^{(n')}GD^{(n)}G](t_{1}-t_{2})$ 
(We write $[AB](t-t')\equiv\int_{C}dt''A(t-t'')B(t''-t')$ and 
subscripts are partially suppressed).
Important cooperon behavior (eq. (\ref{eq:Cdef})) arises in 
$Q_{\kv}^{{\rm (a)}<}$ only when 
all $G_{\kv_{i}}$'s in $D^{(n)}_{ \{\kv_{i}\}, \omega_{1}-\omega }$ 
and $D^{(n')}_{ \{\kv_{i}'\}, \omega_{1}-\omega'}$ 
are retarded and $G_{-\kv_{i}+\pv}$'s in 
$D^{(n)}_{ \{ -\kv_{i}+\pv \},\omega_{1}-\omega-\omega'}$ and
$D^{(n')}_{ \{ -\kv_{i}'+\pv \},\omega_{1}-\omega-\omega'}$ are advanced Green 
functions and for $p\sim0$. By use of 
\begin{equation}
    \sum_{n=0}^{\infty} \sum_{\kv_{i}} 
    {D^{(n) r}_{ \{\kv_{i}\},\omega_{1}-\omega }}
    {D^{(n) a}_{ \{ -\kv_{i}+\pv \},\omega_{1}-\omega-\omega'} }
    \simeq C_{\pv,\omega'}\;\;\;\;\;(p\sim0)
    \label{eq:Cpomega}
\end{equation}
where $C_{p\omega}\equiv 1/[(Dp^2+1/\tau_{\varphi}-i\omega)\tau]$, 
dominant contribution of (\ref{eq:QcoopSE}) is calculated as
\begin{eqnarray}
    Q_{\kv}^{{\rm (a)}<} (t,t,\omega)
        & = & \sum_{\omega' \omega_{1}} 
      e^{-i\omega' t} 
      (\nimp\vi^2)^2 
      (f_{\omega_{1}-\omega-\omega'}-f_{\omega_{1}-\omega'}) 
      [G D^{(n)} G E G D^{(n')} G]^{r}_{\omega_{1}-\omega'}
     [ G D^{(n)} G D^{(n')} G]^{a}_{\omega_{1}-\omega-\omega'}
     \nonumber\\
    & = & \sum_{\omega' \omega_{1}} 
      e^{-i\omega' t} 
      (\nimp\vi^2)^2 
      (f_{\omega_{1}-\omega-\omega'}-f_{\omega_{1}-\omega'}) 
      \sum_{\pv} C_{\pv\omega}C_{\pv\omega'}
       \nonumber\\
    && \times 
       \sum_{\kv'}
      G_{\kv,\omega_{1}-\omega}^r  G_{\kv',\omega_{1}-\omega}^r 
      E_{\omega_{1}-\omega,\omega_{1}-\omega'}^r
      G_{\kv',\omega_{1}-\omega'}^r G_{-\kv+\pv,\omega_{1}-\omega'}^r
       \nonumber\\
    && \times 
      G_{-\kv+\pv,\omega_{1}-\omega-\omega'}^a
      G_{-\kv'+\pv,\omega_{1}-\omega-\omega'}^a 
      G_{\kv,\omega_{1}-\omega-\omega'}^a
     \label{eq:QcoopSE2}
\end{eqnarray}
Retarded part of $E(t,t')$ is given as
\begin{eqnarray}
        E^r (t,t')&=&(\nimp\vi^2)^2 \sum_{nn'=0}^{\infty} 
	\int_{-\infty}^{\infty} dt_{1} \int_{-\infty}^{\infty}  dt_{2}
    [{D^{(n)}}^r G^r](t-t_{1}) 
    \nonumber\\
    &&\times
    i[ \chi^<(t_{1},t_{2}) {F^{nn'}}^r(t_{1}-t_{2}) 
      +\chi^r(t_{1},t_{2})  {F^{nn'}}^>(t_{1}-t_{2})]
    [G^r {D^{(n')}}^r](t_{2}-t')
    \label{eq:Er}
\end{eqnarray}
In terms of Fourier transform (Fig. \ref{FIGAB_AASKL}),
\begin{eqnarray}
   E_{\omega_{1}-\omega,\omega_{1}-\omega'}^r  & = &  
   \sum_{\omega_{4}}
    [{D^{(n)}}^r G^r]_{\omega_{1}-\omega}
    i[ \chi^r_{\omega_{1}-\omega_{4}-\omega,\omega_{1}-\omega_{4}-\omega'} 
    {F^{nn'}}^>_{\omega_{4}} +\chi^< {F^{nn'}}^r]
    [G^r {D^{(n')}}^r]_{\omega_{1}-\omega'}
    \nonumber  \\
     & \simeq &  -i(\nimp\vi^2)^2\sum_{\omega_{4}} (1-f_{\omega_{4}})
    \chi^r_{\omega_{1}-\omega_{4}-\omega,\omega_{1}-\omega_{4}-\omega'} 
    [{D^{(n)}}^r G^r]_{\omega_{1}-\omega}
    [GD^{(n')}GD^{(n)}G]_{\omega_{4}}^a 
    [G^r {D^{(n')}}^r]_{\omega_{1}-\omega'}
    \nonumber \\
     & = &  -i(\nimp\vi^2)^2\sum_{\omega_{4}} (1-f_{\omega_{4}})
    \chi^r_{\omega_{1}-\omega_{4}-\omega,\omega_{1}-\omega_{4}-\omega'} 
    \sum_{\pv'} C_{\pv',\omega_{1}-\omega_{4}-\omega}
      C_{\pv',\omega_{1}-\omega_{4}-\omega'} 
    \nonumber\\
    &&\times
    \sum_{\kv_{1}\kv_{2}}  G_{\kv_{1},\omega_{1}-\omega}^r  
    G_{\kv_{1}-\Qv,\omega_{4}}^a  G_{-\kv'+\pv',\omega_{4}}^a   
    G_{\kv_{2}-\Qv,\omega_{4}}^a  G_{\kv_{2},\omega_{1}-\omega'}^r
    \label{eq:Eromega}
\end{eqnarray}
We thus obtain
\begin{eqnarray}
    Q_{\kv}^{{\rm (a)}<} (t,t,\omega)
        & = &  -i\nimp\vi^2 \tau^2 \sum_{\omega' \omega_{1}\omega_{4}}
	      e^{-i\omega' t} \sum_{\pv}
 G_{\kv,\omega_{1}-\omega}^r  G_{-\kv+\pv,\omega_{1}-\omega'}^r
 G_{-\kv+\pv,\omega_{1}-\omega-\omega'}^a G_{\kv,\omega_{1}-\omega-\omega'}^a
       \nonumber\\
      && \times
         (f_{\omega_{1}-\omega-\omega'}-f_{\omega_{1}-\omega'}) 
     (1-f_{\omega_{4}})
\chi^r_{\omega_{1}-\omega_{4}-\omega,\omega_{1}-\omega_{4}-\omega'} 
         C_{\pv\omega}C_{\pv\omega'}  
      \sum_{\pv'} C_{\pv',\omega_{1}-\omega_{4}-\omega}
      C_{\pv',\omega_{1}-\omega_{4}-\omega'} 
            \nonumber\\
      && \times
      (2+3i\tau(\omega_{1}-\omega_{4})-4p^2 D\tau)
\end{eqnarray}
Other processes in Fig. \ref{FIGAB_AAS} (b)(c) are similarly calculated as
\begin{eqnarray}
    Q_{\kv}^{{\rm (b+c)}<} (t,t,\omega)
        & = & -i\nimp\vi^2 \tau^2 \sum_{\omega' \omega_{1}\omega_{4}} 
      e^{-i\omega' t} 
      \sum_{\pv}
 G_{\kv,\omega_{1}-\omega}^r  G_{-\kv+\pv,\omega_{1}-\omega'}^r
 G_{-\kv+\pv,\omega_{1}-\omega-\omega'}^a G_{\kv,\omega_{1}-\omega-\omega'}^a
      \nonumber\\
      && \times
      (f_{\omega_{1}-\omega-\omega'}-f_{\omega_{1}-\omega'}) 
      (1-f_{\omega_{4}})
    \chi^r_{\omega_{1}-\omega_{4}-\omega,\omega_{1}-\omega_{4}-\omega'} 
      C_{\pv\omega}C_{\pv\omega'}  
      \sum_{\pv'} C_{\pv',\omega_{1}-\omega_{4}-\omega}
      C_{\pv',\omega_{1}-\omega_{4}-\omega'} \nonumber\\
      &&  \times
     (-2-i\tau(4\omega_{1}-4\omega_{4}-\omega-\omega')+5p^2 D\tau)
\end{eqnarray}
It is seen that one of the four cooperons is canceled after summation 
of the three processes (a-c)\cite{Fukuyama85}, and we obtain 
$Q_{\kv}^{{\rm AAS}<}\equiv Q_{\kv}^{{\rm (a)}<}+Q_{\kv}^{{\rm (b+c)}<} $ as
(noting $p,p'\ll k$ and $\omega'\tau\ll 1$)
\begin{equation}
    Q_{\kv}^{{\rm AAS}<}(t,\omega\rightarrow0)=-i\omega \nimp\vi^2 \tau^2  
     \sum_{\omega' \omega_{1}} e^{-i\omega' t} 
      (G_{\kv}^r  G_{\kv}^a)^2 
      \sum_{pp'} 
     C_{p\omega} C_{p\omega'} C_{p'\omega_{1}}       
       \left(f_{\omega_{1}-\omega'}\chi^r_{\omega_{1},\omega_{1}-\omega'} 
        - f_{\omega_{1}}\chi^a_{\omega_{1},\omega_{1}-\omega'} \right)
    \label{eq:QAAS}
\end{equation}
where $\chi^a$ term is due to the complex processes 
(Fig. \ref{FIGAB_AAS}(d)) and $G_{\kv}\equiv G_{\kv,\omega=0}$.

\begin{figure}[bthp]
\begin{center}
\epsfxsize=10cm              
\epsfbox{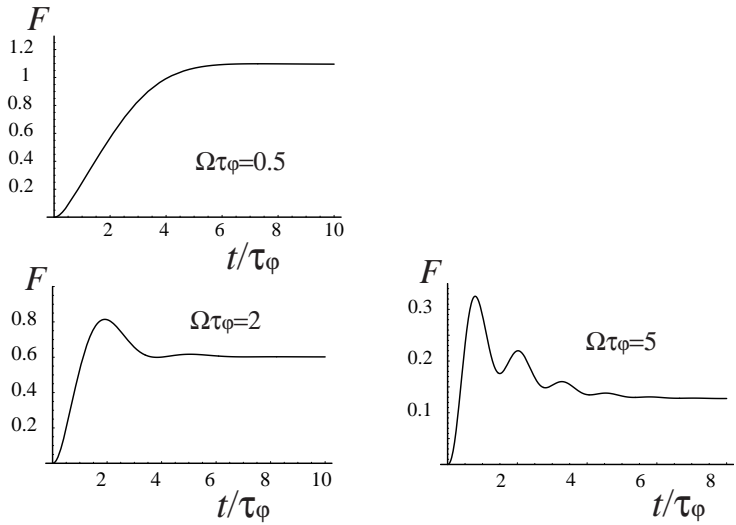}
\end{center}
\caption{
Relaxation of the amplitude of AAS oscillation ($F$ of eq. 
(\ref{Fdef})) after switching of TLS for $\Omega\tauphi=0.5,2,5$.
For $\Omega\tauphi\ll 1$ behavior is monotonic, but for 
$\Omega\tauphi\protect\agt1$, a bump appears in the 
timescale of $\sim\tauphi$ and then a decay.
\label{FIGAB_AASc}
}
\end{figure}

The current at low temperature is obtained by use of eq. (\ref{eq:chisft}) 
(and (\ref{eq:J0}) ) as
\begin{eqnarray}
  \delta J^{\rm AAS}(t)&=& J_{0}\frac{\pi v^2}{m k_{F} V} 
    \sum_{\pv\pv'}\frac{1}{\Ap} \int_{-\infty}^{0} d\omega \sum_{\pm} (\pm)
    \nonumber\\
    && \times
    {\rm Im} 
    \left[ -\left(1-e^{-\Ap t} -i\Ap e^{-\Ap t}
    \frac{e^{-i(\omega\pm\Omega)t}-1}{\omega\pm\Omega} \right) 
    \frac{3\omega\pm\Omega+i(2\App-\Ap)} 
    {[(\omega\pm\Omega)^2+\Ap^2] (\omega+i\App) [\omega+i(\App-\Ap)]}
    \right.
    \nonumber\\
    &&
    +\left. \left(\frac{1-e^{-(\App-i\omega)t}}{\omega+i\App} 
    -e^{-(\App-i\omega)t} 
    \frac{e^{-i(\omega\pm\Omega)t}-1}{\omega\pm\Omega}\right) 
    \frac{1}{[\omega+i(\App-\Ap)](\pm\Omega-i\App)} \right]
\end{eqnarray}
where $\Ap\equiv{Dp^2+1/\tau_{\varphi}}$.
Slowest relaxation is governed by the contribution from $p=p'=0$ of the 
square bracket part.
The oscillation part of this contribution is 
obtained as
\begin{equation}
 \delta J^{\rm AAS}(t) \simeq  J_{0} \cos(2\phase)  \frac{3}{4\pi} 
    \frac{\ell_{\varphi}}{\ell}\frac{v^2}{k_{F}^2 ab} \tauphi^3 F
\end{equation}
where
\begin{eqnarray}
    F&\equiv&   {\rm Im} \int_{-\infty}^{0} d\omega \sum_{\pm} 
    \frac{\pm}{\tauphi^3}
    \left[ -\left(1-e^{- t/\tauphi} -i\frac{e^{- t/\tauphi}}{\tauphi} 
    \frac{e^{-i(\omega\pm\Omega)t}-1}{\omega\pm\Omega} \right) 
    \frac{3\omega\pm\Omega+i/\tauphi} 
    {[(\omega\pm\Omega)^2+\tauphi^{-2}] (\omega+i/\tauphi) \omega}
    \right.
    \nonumber\\
    &&
    +\left. \left(\frac{1-e^{-(1/\tauphi-i\omega)t}}{\omega+i/\tauphi} 
    -e^{-(1/\tauphi-i\omega)t} 
    \frac{e^{-i(\omega\pm\Omega)t}-1}{\omega\pm\Omega}\right) 
    \frac{1}{\omega(\pm\Omega-i/\tauphi)} \right]
    \label{Fdef}
\end{eqnarray}
Enhancement of AAS current ($\delta J^{AAS}>0$) is explained as due to
the dephasing effect of TLS, which suppresses localization. 
The phase of the oscillation is not modified (i.e., $\delta 
J^{AAS}\propto \cos2\phi$), and only the amplitude relaxes after TLS 
is switched.
If $\Omega \ll \tauphi$ the time scale of the relaxation is $\sim 
\Omega^{-1}$. In the opposite case of $\Omega \gg \tauphi$, there 
appears first a rise in the timescale of $\tauphi$ 
followed by a rapid decay with small oscillation of 
frequency of $\sim \Omega$ (Fig. \ref{FIGAB_AASc}).
The effect of TLS vanishes as $\propto \Omega$ for $\Omega\ll1$ and as 
$\propto 1/\Omega$ for $\Omega\gg 1$.
Vanishing of the effect in these limits is different from the 
ballistic case (\ref{eq:JAB4}) and consistent with the 
explanation by dephasing effect. 


\section{Summary and Discussion}
We have calculated the electronic current through an Aharonov-Bohm (AB)
ring after a quantum Two-Level-System (TLS) is switched on.
TLS affects the amplitudes of 
AB and AAS oscillations, which relax to new equilibrium values. 
Phases of both oscillations are not affected.
If the energy splitting of the TLS, $\Omega$, is small,
the time scale of the amplitude relaxation is given  by
the characteristic time of the system, which is elastic lifetime 
$\tau$ in the case of ballistic and inelastic lifetime $\tauphi$ in 
diffusive case.
In the opposite case, $\Omega \ll \tau,\tauphi$, the time scale becomes 
$\Omega^{-1}$.
Although the relaxation of current appears similar in both ballistic and 
diffusive case, physics behind is different.
In the ballistic case the relaxation is due to a scattering of the 
states into other states, which is not dephasing. In the 
diffusive case the relaxation is interpreted as due to dephasing. 
Crucial difference  between the two is that in the diffusive case, 
one one hand, phase produced 
by TLS is randomly accumulated because of contribution from random 
paths the electron travels, while in the ballistic case on the other hand, 
there is no randomness. 
Dephasing effect would appear in ballistic case if the energy of the TLS is 
distributed. 

Effect of oscillating external field is also calculated.
The amplitude of the current oscillates if the external oscillation is 
slow enough for the electron to accommodate, but current oscillation 
is not obvious in the fast varying case.

Recent high (THz) time-resolved measurements of electronic 
properties\cite{Beard00} would make it possible to 
observe the current response and time-resolved dephasing processes. 
The current response may provide us direct information on microscopic 
relaxation times (elastic ($\tau$) and inelastic lifetime 
($\tauphi$)) and properties of the perturbation source.

Current measurement may be a useful tool in mesoscopic dynamics. 
For instance, a motion such as 
slow oscillation of a magnetic domain wall in 
nano-scale magnetic contacts\cite{Garcia99,Versluijs01} may be 
detectable as an oscillation of electronic current through the contact. 
Time-resolved transport measurement may become a new and powerful method
in studying mesoscopic dynamics.

\acknowledgements
\noindent

G.T. thanks H. Matsukawa and H. Kohno for valuable discussion. 
He is grateful to The Mitsubishi Foundation  for financial support.


%
\end{document}